\documentclass[11pt,a4paper]{article}
\usepackage[utf8]{inputenc}
\usepackage{geometry}
\usepackage{amsfonts}
\usepackage[toc]{appendix}
\usepackage{xcolor}

\usepackage{physics}
\usepackage{amsmath}
\usepackage{tikz}
\usepackage{mathdots}
\usepackage{yhmath}
\usepackage{cancel}
\usepackage{mathtools}
\usepackage{color}
\usepackage{siunitx}
\usepackage{array}
\usepackage{multirow}
\usepackage{amssymb}
\usepackage{gensymb}
\usepackage{tabularx}
\usepackage{extarrows}
\usepackage{booktabs}
\usetikzlibrary{fadings}
\usetikzlibrary{patterns}
\usetikzlibrary{shadows.blur}
\usetikzlibrary{shapes}

\usepackage{dsfont}

\usepackage{cite}
\usepackage{graphicx}
\usepackage{braket}
\usepackage{slashed}
\usepackage{float}
\usepackage{braket}

\usepackage{upgreek}
\usepackage{hyperref}

\hypersetup{
	colorlinks,
	linkcolor={blue!50!red},
	citecolor={orange!70!red},
	urlcolor={orange!20!red}
} 

	\def\A{\mathcal{A}}

\begin{document}
		\def\stateob{\Psi_{\rm obs}}
		\def\trace{{\rm Tr}}
		\def\algebra{\hat{\A}_{\rm cr}}
		\def\HH{\Psi_{\rm HH}}

		\begin{titlepage}
			\begin{flushright}
				\small{\it Version Dated: \today}
			\end{flushright}
			\vskip 1.5in
			\begin{center}
				
				{{{\LARGE \textsc{ The Canvas of Holography in}  (A)d\textsc{S/CFT}}}}
				\vskip
				0.7cm {Vaibhav Kalvakota and Aayush Verma} 
				\vskip 0.4in {\small{\vskip -1cm
						{Email: 
        \texttt{vaibhavkalvakota@gmail.com}, \texttt{aayushverma6380@gmail.com}}}}
			\end{center}
			\vskip 0.5in
			\baselineskip 16pt
			\begin{abstract}

    The dynamic of holography between anti-de Sitter space holography and de Sitter holography is a very fascinating comparison, which provides many key insights into what we expect from holography in general. In this Essay, we highlight this dynamic with three examples: first, when taking Wheeler-DeWitt states to the asymptotic boundary, the dual interpretation is unclear in de Sitter. Second, what we make of bulk reconstruction and subregion duality in AdS/CFT is not trivially reflected in the dS/CFT scenario. Third, a way of formulating emergence and subregion-subalgebra duality in de Sitter space does not yet exist. With these examples, we provide some musings on this canvas of holography in the settings of (A)dS/CFT.
    
			\end{abstract}
			
			\vskip 1in
			\begin{flushleft}
				\textit{Essay written for the Gravity Research Foundation 2024 Awards for Essays on Gravitation} \end{flushleft}
		\end{titlepage}



\newpage
		
		{\hypersetup{linkcolor=black}

		\textbf{Introduction:} AdS/CFT was first described by Maldacena as a 10d type IIB string theory compactified on AdS$_{5}\times S^{5}$ dual to a $\mathcal{N}=4$ super Yang-Mills conformal field theory on the 4d boundary of the bulk theory \cite{Maldacena:1997re, Witten:1998qj,Aharony:1999ti}. The number of emergent dimensions from boundary to bulk is six here. However, this definition is in the context of string theory and is certainly a hallmark discovery of string theory. That definition is very specific and can be boiled down into a more general mantra by saying that the CFT partition function $Z_{CFT}$ is equivalent to the generating functional for the bulk theory:
\[Z_{CFT}=Z_{bulk}\;.\]

While this is a general enough statement for (in principle) {\it any} holographic theory, there are two issues with this statement: (1) the theory dual to a bulk theory need not be a \textit{conformal field theory} at all and could be a general QFT, which is the deformation of some CFT, as is the case with finite cutoff holography \cite{McGough:2016lol,Shyam:2018sro,Araujo-Regado:2022gvw}, and (2) this definition does not explain the localization of information in the sense which gravity does and the gravity is special for its role in localizing information at infinity \cite{Laddha:2020kvp,deMelloKoch:2022sul}. We will discuss this later in this essay. What we are trying to point out is that the definition of ``holography" is as mysterious as the theory itself, and there is no {\bf one} obvious way of defining what holography really is. Moreover, throughout this essay, our definition would be traditional while there have been many modern re-definitions of the term, and giving an all-round literature reference is, of course, not possible. Additionally, this paper does not act as a literature review.

Firstly, there are many reasons why we do holography and \textit{how} we do holography. One could be interested in doing bulk reconstruction, and for instance, taking bulk fields $\phi _{bulk}$ to the asymptotic limit $\rho \to \infty $ (where $\rho $ describes the variation of fields towards the boundary) and identify an HKLL-style scheme, which is a very well-understood thing in AdS/CFT. Or one could expand on this a little more and be worried with bulk reconstruction in the entanglement wedge $\mathcal{E}_{W}(R)$ of some boundary subregion $R$, which is also a very well understood problem. In this particular case, we are explicitly dealing with the information side of duality; we have much more convenience with how we use holography. On the other hand, we have recently obtained algebraic interpretations like subalgebra-subalgebra correspondence \cite{Leutheusser:2021frk,Leutheusser:2022bgi}, which we will discuss later. 

The best example of why the usual description of holography is a little vague in some cases is that of the famous conundrum of de Sitter holography at $\mathcal{I}^{\pm }$ -- referred to as dS/CFT. In this case, trying to do things asymptotically from the bulk picture is well-defined, such as calculating S-Matrix at infinity. Still, it is not clear what exactly the \textit{holographic} interpretation is in dS/CFT. In AdS/CFT, one could start from asymptotic canonical quantum gravity, where we first try to solve the Wheeler-DeWitt equation in the radial limit, so that the WDW states thus obtained become CFT partition functions like\footnote{Here, $g_{r}$ represents the original AdS metric. Usually, one rescales the metric by some parameter $\mu $ which plays the role of the deformation parameter in doing $T\overline{T}$-deformations. More generally, the rescaled metric is what the WDW equation is solved asymptotically with, and the matter fields $\Phi $ are also appropriately rescaled, which we use next. Here we are abusing terminology for the sake of brevity.}
\begin{equation}\label{wdw}
    \lim _{r \to \infty }\Psi [g_{r}, \Phi ]=Z^{\pm }[g_{r}, \Phi ]
\end{equation}
The field theory dual to this is clear enough in AdS/CFT, but in de Sitter this is not the case. Due to this, making sense of the $Z^{\pm }$ branches as the field theory dual is unclear. In Eq.~\eqref{wdw}, the duality between WDW state and partition function is a special deduction of the holography, as propounded in \cite{Raju:2019qjq}. In such cases, we have an obvious understanding of some holography of information, if not holography in itself, of which \cite{Chakraborty:2023los} is a good example; also see \cite{dsnote} for a review.

\textbf{Wheeler-DeWitt States:} In de Sitter, for our purposes, we will now refer to the metric in use as $\widetilde{g}_{\mu \nu }$, which is described in terms of the original de Sitter metric 
		\begin{equation}
			\widetilde{g}_{\mu \nu }=\rho ^{2}g_{\mu \nu }\;,
		\end{equation}
		so that the asymptotic boundary is the limit $\rho \to 0$. The matter fields are also described according to the rescaled metric, denoted as $\widetilde{\Phi }$, and similarly, in $\rho \to 0$, the matter fields are taken to the boundary. The WDW state $\Psi [\widetilde{g}, \widetilde{\Phi }]$ is then the solution to the WDW constraint equation,
		\begin{equation}\label{WDEEQ}
			\hat{H}\Psi [\widetilde{g}, \widetilde{\Phi }]=0\;,
		\end{equation}
		and when taking these bulk states to the asymptotic boundary, we recover things that \textit{look} like the partition functions to the boundary theory, similar to Eq.~\eqref{wdw}. 
  
        The solutions to the Eq.~\eqref{WDEEQ} are the Wheeler-de Witt states, which are written as
		\begin{equation}\label{branch}
			\lim _{\rho \to 0} \Psi [\widetilde{g}, \widetilde{\Phi }]=e^{i\mathcal{S}[\widetilde{g}, \widetilde{\Phi }]}Z^{+}[\widetilde{g}, \widetilde{\Phi }]+e^{-i\mathcal{S}[\widetilde{g}, \widetilde{\Phi }]}Z^{-}[\widetilde{g}, \widetilde{\Phi }]\;.
		\end{equation}
        which is a general solution to the Hamiltonian constraint equation. In de Sitter's case, $e^{iS}$ is just a phase factor, and $Z^\pm$ are CFT-like functionals. We will discuss this more in a moment.
		
		In this sense, the Hilbert space is constructed out of the analytically continued Euclidean vacuum state $|0\rangle $; see \cite{Chakraborty:2023yed} for details. That is, we can write something like 
		\begin{equation}
			|\Psi \rangle =\int dx_{1}\dots dx_{n} \psi (x_{1}\dots x_{n})\Phi (x_{1}\dots x_{n})|0\rangle \;,
		\end{equation}
		where for $n>0$, this usually implies that the states are not invariant under the $dS_{D}$ isometry group $SO(D,1)$, with the exception of the vacuum state itself. Furthermore, they have infinite norms and are not normalizable. So far, this is only producing the \textit{auxiliary} Hilbert space $\mathcal{H}_{aux}$, and the constraint to be imposed is that the states $|\Psi _{i}\rangle \in \mathcal{H}_{aux}$ are invariant under the dS isometry group $G_{dS}$, which form the \textit{physical} Hilbert space $\mathcal{H}_{phys}$.  Following Moncrief's conjecture \cite{moncrief1976space} and Higuchi's construction \cite{higuchi1991quantum,higuchi1991quantum2}, we can group average by taking the norm and dividing it by the volume of the Diff$\times$Weyl group $\mathrm{vol}\left(\text{Diff}\times \text{Weyl} \right)$, so as to produce a finite norm for the states, and $\mathcal{H}_{phys}$. This is in the $\kappa \to 0$ limit, where we obtain the modified norm and the physical Hilbert space, and these states are invariant, as required by the Gauss law. For a BRST interpretation of these states, see \cite{Ljatifi:2023dro}. 
		
		$Z^{\pm }[\widetilde{g}, \widetilde{\Phi }]$ are CFT-like functionals, in that they obey the conformal Ward identity and have the same properties such as diffeomorphism invariance and Weyl transformation properties. In $D=2$, the anomaly obtained $\A_D$ is the usual central charge, and is just the analytically continued Brown-Henneaux central charge. In the AdS/CFT context, this is very clear since one can look at the asymptotic $Z^{\pm}[\widetilde{g}, \widetilde{\Phi }]$ as a deformed theory in the $\rho \to 0$ limit, which acts as a deformation parameter. In de Sitter, however, the holographic dual to these WDW states is not very clear.
		
		Let us look a little more clearly into the constraints that are being imposed on $\Psi [\widetilde{g}, \widetilde{\Phi }]$ and $Z^\pm [\widetilde{g}, \widetilde{\Phi }]$, which are very important. First, we expect that the WDW states are invariant under the action of $G_{dS}$ isometry group, for which we have to modify the definition of the states by using the group-averaging construction by Moncrief.  On the other hand, the CFT-like generating functionals for the field theory would obey diffeomorphism invariance and the conformal Ward identity, where the latter can be expressed into anomaly terms that look like 
		\begin{equation}
			\mathcal{A}_{D}Z[\widetilde{g}, \widetilde{\Phi }]\;.
		\end{equation}
    where when $D$ is even $\A_D$ can be computed and otherwise when $D$ is odd, it vanishes.
		One would expect that in the perturbative setup, the asymptotic $Z$-functionals obey a Weyl anomaly property with an extra imaginary term, due to which in even $D$ there would be an anomaly term with an extra factor of $i$ than in AdS \cite{Chakraborty:2023yed}. 
		
		The general idea here is that when trying to interpret holography in the canonical quantum gravity formalism, there is a natural link to the holographic renormalization counterterms arising from deforming the theory. In AdS/CFT, this deformation would give us finite-cutoff holography, but in dS this is not a very obvious setting. In deriving finite-cutoff AdS holography from deformations, one can see that in the limit of $\rho \to 0$, one of the branches is ``dominant", due to which the WDW states collapse to
		\begin{equation}
			\Psi [\widetilde{g}, \widetilde{\Phi }]=e^{iS[\widetilde{g}, \widetilde{\Phi }]}Z[\widetilde{g}, \widetilde{\Phi }]\;.
		\end{equation}
		The term $S[\widetilde{g}, \widetilde{\Phi }]$ acts as a holographic renormalization term and more appropriately in the de Sitter setup is a local action that is universally factored to the WDW states\footnote{In the perspective of deformations these are counterterms in doing irrelevant deformations, given by the $T\overline{T}$ (or $T^{2}$-deformations in $D>2$) deformations \cite{Araujo-Regado:2022gvw}.} \cite{Chakraborty:2023los}. In dS, there is no one dominant branch, due to which both branches in Eq.~\eqref{branch} are taken into account. In describing WDW states in the large-volume limit, we set $\kappa =\sqrt{8\pi G_{N}}\to 0$ limit, in which the states look similar to the above single-$Z$ form.  In the static patch picture, one would use $T\overline{T}+\Lambda _{2}$ deformations\footnote{The bulk 3D case is more relevant in recent constructions by Vasu Shyam, Eva Silverstein, etc., due to which the 2D CFT case is best suited for deformations, hence the $\Lambda _{2}$ term rather than the more general $\Lambda _{D-1}$.}, although interpreting these in the canonical QG sense and approaching Cauchy slice holography is still an open problem. To give a more general overview of why these WDW states are of importance, the holographic context is the most interesting background for working on problems in (A)dS. Naturally, these WDW states in a holographic sense are dual to the partition function of the dual CFT:
		\begin{equation}
			\Psi [\widetilde{g}, \widetilde{\Phi }]=Z_{\text{CFT}}[\widetilde{g}, \widetilde{\Phi }]\;.
		\end{equation}
		This statement can be found backwards by taking the asymptotic limit, as discussed above. In de Sitter, we expect that there are two CFT-like functionals, $Z^{+}[\widetilde{g}, \widetilde{\Phi }]$ and $Z^{-}[\widetilde{g}, \widetilde{\Phi }]$, which seem straightforward seeing that there are \textit{two} asymptotic boundaries towards $\mathcal{I}^{+}$ and $\mathcal{I}^{-}$, and the corresponding anomaly terms have different signs. However, this is also expected more generally in AdS, since the Hamiltonian is quadratic in nature with $Z[\widetilde{g}, \widetilde{\Phi }]$ being linear, implying two such terms. 
		
\textbf{Subregion Duality:} Subregion duality for de Sitter space is a much more controversial situation. One of the reasons why subregion duality is not entirely obvious (or concrete, for that matter) in de Sitter holography is because in the traditional candidate for de Sitter holography, dS/CFT, the expectation of finding subregion duality from a Ryu-Takayanagi-like formula is not clear. One of the ways of looking at why subregion duality is complicated in dS/CFT is to take inspiration from the usual AdS/CFT setting.
		
		First, we will offer a perspective on why modular flows naturally arise in AdS/CFT. From the extrapolate dictionary\footnote{In dS/CFT, the extrapolate dictionary is different from the \textit{differentiate} dictionary, for which there are two conformal weights, for which there would exist a double reconstruction scheme for $\mathcal{I}^{\pm }$ respectively.},
		\begin{equation}
			\mathcal{O}_{\Phi }\xrightarrow[r\to \infty ]{\Delta } \Phi (\mathbf{Y}) \;,
		\end{equation}
	where $\mathbf{Y}$ is the bulk point. From this, one could define a spacelike region formed by the overlap of the double-lightcone $\mathcal{L}(\mathbf{Y})$ and use the HKLL construction for reconstructing $\Phi (\mathbf{Y})$ dual to $\mathcal{O}_{\Phi }$. In dS/CFT, this can be done up to an extent since there exists an HKLL-type reconstruction scheme for taking bulk fields to the asymptotic boundaries $\mathcal{I}^{\pm }$ \cite{Xiao:2014uea}. However, one does not have an obvious notion of an entanglement wedge associated to boundary subregions. The natural motivation would be from the holographic entanglement entropy prescription
	\begin{equation}
		S(R)=\frac{\text{Area of }\gamma _{R}}{4G_{N}}+S_{\text{bulk}}(r)\;,
	\end{equation}
where $R$ is a boundary subregion, $\gamma _{R}$ is the corresponding Ryu-Takayanagi surface and $r$ is the bulk subregion dual to $R$. The domain of dependence $\mathcal{D}(\gamma_R)$ of $\gamma_R$ is the entanglement wedge $\mathcal{E}_{W}(R)$ so that bulk fields in $\mathcal{E}_{W}(R)$ can be described by dual Heisenberg single-trace operators in $R$. This is a standard known result. A natural consequence was that the relative entropies are equivalent for these subregions, giving the JLMS formula \cite{Jafferis:2015del,Faulkner:2017vdd,Dong:2016eik}: 
\begin{equation}
	S(\rho |\sigma )=S_{\text{bulk}}(\rho |\sigma )\;.
\end{equation}
This result is found from the machinery of modular flows, but in dS a clear problem is that taking coordinates $(\eta , \chi )$ with $\eta $ flowing from $\mathcal{I}^{+}$ to $\mathcal{I}^{-}$, a variation of $\eta $ is also a variation from the boundary $\partial M$, which is not the case for AdS -- calling the coordinate counterparts in AdS $(\tau , r )$, a variation in $r$ is not a variation from $\partial M$ since the boundary $\partial M$ lies on $r\to \infty $ slices along $\partial \Sigma $. In this sense, modular flows are not trivial in de Sitter. 

One naive possibility arises from the existence of pseudo-quantities from the timelike entanglement entropy \cite{Doi:2022iyj,Doi:2023zaf}, from which one could argue that instead of dealing with density matrices for modular flows like 
\begin{equation}\label{eq:adsmod}
	\rho ^{-is}\mathcal{O}\rho ^{is}\;,
\end{equation}
we could deal with the transition matrices $\widetilde{\rho }$ to find something like 
\begin{equation}
	\widetilde{\rho }^{~-is}\mathcal{O}\widetilde{\rho }^{~is}\;.
\end{equation}
However, there are two very obvious subtleties with this. First, the role of the modular parameter $s$ is akin to the time $\tau $, in which case we are essentially taking local boundary operators $\mathcal{O}$ and performing modular flows labelled by $\tau \equiv s$. The role of the entanglement wedge is then natural from the reconstruction scheme for $\mathcal{O}_{R}\in \mathcal{D}(R)$. So, the role of the modular parameter in dS/CFT changes and is not entirely clear. Second, even if there does exist some similar interpretation to \eqref{eq:adsmod} in dS, whether that necessarily defines a JLMS-type formula is non-trivial since that would require defining the action of the modular Hamiltonian and the corresponding notion of the code subspace. For these reasons, interpreting subregion duality with the same tone as in AdS/CFT in dS/CFT is very difficult to speculate on. There have been several works providing a more geometric construct for subregion duality, which are based on the Wick rotation scheme \cite{Narayan:2020nsc,Narayan:2023zen,Narayan:2017xca}, but to say whether these have any clear association to the above-mentioned subtleties is non-trivial. 

Another candidate for working with modular flows and possibly subregion duality is that of static patch holography, where the boundary (the cosmic horizon) does not run along a timelike coordinate such as $\eta $, due to which it is a suitable setting for working with modular flows\footnote{We would like to thank Vasu Shyam for discussions on this.}. Static patch holography is also an interesting setting for working with $T\overline{T}+\Lambda _{2}$ deformations, on which a lot of interest has been sparked recently \cite{Coleman:2021nor,Lewkowycz:2019xse,Batra:2024kjl,Shyam:2021ciy}. In static patch de Sitter, one can also have a notion of entanglement wedge from the monolayer and bilayer proposals as pointed out in \cite{Franken:2023pni,Franken:2024ruw} \footnote{These proposals find an origin from the bit threads formalism of calculating entanglement entropy as first formulated in \cite{Shaghoulian:2022fop}, through which an RT-like formula can be found using a non-zero extremal surface lying between the stretched horizons homologous to the podes. We thank Victor Franken for pointing out these references.}.

\textbf{Operator Algebras:} One does not necessarily have to work with the information side of things, and instead be interested in working with the algebraic description of the duality. In that sense, AdS/CFT boils down to subregion-subalgebra duality \cite{Leutheusser:2022bgi} (as opposed to subregion-subregion duality, which solely tells us what the dual subregions in the bulk and boundary are and how the entropies are related), which is the statement that the Hilbert space of the bulk theory, which is the asymptotic Fock space, is equivalent to the GNS-constructed Hilbert space from the boundary side:
\begin{equation}
    \mathcal{H}_{bulk}=\mathcal{H}_{GNS}\;.
\end{equation}
In this sense, the Liu-Leutheusser result of the emergence of bulk locality tells us that the emergent type III$_{1}$ algebra on the boundary gives us the emergent bulk physics \cite{Leutheusser:2021frk,Leutheusser:2022bgi}. Taking the large $N$ algebra and adding the renormalized Hamiltonian to it, one can do the crossed product construction due to Chandrasekaran, Pennington, and Witten \cite{Chandrasekaran:2022eqq,Witten:2021unn}, which yields a type II$_{\infty }$ algebra, for which there exists a trace class for which a trace can be defined. This factor describes the algebra of observables for black hole states \cite{Witten:2021unn}. The crossed product is taken between the factor type III$_1$ and its modular automorphism group. One can then attribute entropy to states in subalgebras $\mathcal{B}\subseteq \mathcal{A}$ (where $\mathcal{A}$ is the full boundary von Neumann algebra) whose bulk description is the generalized entropy prescription. That is, the entropy $\mathcal{S}_{E}(\Psi _{\mathcal{B}})$ of $\Psi _{\mathcal{B}}$ is equal to the generalized entropy of some surface in the bulk identified with the QES prescription \cite{Engelhardt:2014gca,Wall:2011hj,Chandrasekaran:2022eqq}. Taking $\mathcal{B}$ to future infinity to get the future infinity subalgebra $\mathcal{B}_{\infty }$, for instance, implies that we are taking the generalized entropy of horizon cuts towards future infinity, and using the monotonicity of trace-preserving inclusions, one can show that the generalized entropy of these horizon cuts to future infinity is greater than the generalized entropy of the bifurcation surface in the two-sided black hole setup. So, we have two ways of looking at holography in the same algebraic context; we could describe the boundary theory in the type III$_{1}$ algebras for which we obtain subregion-subalgebra duality, or do the crossed product construction in the large $N$ limit of the microcanonical ensemble, in which we obtain a type II$_{\infty }$ algebra for which entropies of subalgebras (or more generally \textit{subspaces}) is the generalized entropy of horizon cuts, which lies in the entanglement wedge prescription.  These results were generalized for the algebra of exterior arbitrary black hole spacetimes recently on the basis of the boundedness of the subregion and if it contains an asymptotic boundary \cite{Jensen:2023yxy,Kudler-Flam:2023qfl} \footnote{The differences in \cite{Jensen:2023yxy} and \cite{Kudler-Flam:2023qfl} results were the assumptions taken. See the footnote 5 in \cite{Kudler-Flam:2023qfl}. Moreover, \cite{Jensen:2023yxy} assumes the geometric modular flow conjecture. We thank Gautam Satishchandran for mentioning these points to us.}; see also \cite{AliAhmad:2023etg,Klinger:2023tgi} for related results.

Similarly, we can also construct an algebra of observables for an observer in the de Sitter static patch, first discussed in \cite{Chandrasekaran:2022cip} and see \cite{Witten:2023qsv,dsnote} for reviews. To sum up, in brief, we start with an empty de Sitter static patch, which is but \smallskip type III, and then we include an observer (subject to a Hamiltonian constraint $H_{\text{obs}}\geq 0$) on the worldline with turning the gravity on. This will gravitationally dress operators, and the final algebra for this static patch would be type II$_{1}$. The maximum entropy state is given by the Bunch-Davies state, whose density matrix is $\rho=1$; the entropy of this state is $S=0$. One then obtains the Hilbert space on which these operators act, but as before, states in the Hilbert space are plagued by infinite norms and have to be modified in the same way: to take the norm and divide it by the isometry group of de Sitter. In this way, we get invariant states under the whole isometry group.

The beautiful part of the result on the emergent holography from type III$_{1}$ algebra in AdS/CFT was the subregion-subalgebra correspondence essential to building the emergence and use of half-sided modular inclusion, which is special only to type III$_1$. Alas! We do not have a similar subalgebra-subalgebra correspondence in the de Sitter setting. Even holography in the algebraic sense is different from the geometry in which we work. We would be eager to discover subtleties we miss in de Sitter's algebra of observables in the future.

For some people, holography could be as simple as saying that the information about \textit{stuff} inside some $D$-dimensional system is encoded on the $D-1$-dimensional boundary. Adopting a codimension-1 approach, we would just be left out with holography of information \cite{deMelloKoch:2022sul,Bahiru:2023zlc,Chakraborty:2023los}, which is precise enough in the sense that we know that the bulk in $\Sigma $ is described by a field theory on the boundary $\partial \Sigma $. This has some issues with some aspects of algebraic QFT, specifically that of the split property \cite{Raju:2021lwh}. While there exists a de Sitter counterpart to the AdS holography of information \cite{Chakraborty:2023los}, this is outside the scope of our discussion. 

To sum up, in general we expect holography to consist of a bulk theory $\textsf{R}$ and a boundary theory $\textsf{B}$. This meets three fundamental expectations:
\begin{enumerate}
    \item There exists a way of making sense of holographic entanglement entropy and other quantum information aspects in $\textsf{R}$. We would naturally expect that the prescription for this naively looks similar to the usual Ryu-Takayanagi formula in AdS/CFT, looking in general something like
    \begin{equation}
        S(\textsf{B}_{a})=\text{Area of }{\textsf{R}_{a}}+\text{bulk corrections}\;.
    \end{equation}
    \item There exists a correspondence between the operator algebras associated with $\textsf{R}$ and $\textsf{B}$. Let $\textsf{R}$ be a bulk subregion with a local operator algebra $\mathcal{A}(\textsf{R})$. Dual to this would be a boundary subregion $\textsf{B}$ with the corresponding algebra $\mathcal{A}(\textsf{B})$. Then, the operator algebra on the boundary $\mathcal{A}(\textsf{B})$ and the operator algebra in the bulk $\mathcal{A}(\textsf{R})$ have a relation that should look like
    \begin{equation}
        \mathcal{A}(\textsf{B})\Longleftrightarrow \mathcal{A}(\textsf{R})\;.
    \end{equation}
    Moreover, the complement $\Bar{\textsf{R}}$ has a dual operator algebra of the boundary, so that from the Haag duality, we have
    \begin{equation}
     \mathcal{A}'(\textsf{B})=\mathcal{A}(\Bar{\textsf{B}})\;,
    \end{equation}
    and there exists a possible corresponding bulk reconstruction scheme.
    \item In any holographic theory, there exists a way of solving for the Page curve and reconstructing the interior operators of a black hole in the holographic context so that there exists a way to address the black hole information problem \cite{Almheiri:2019hni,Penington:2019npb,Papadodimas:2012aq}. 
\end{enumerate}

In some cases, we might know more about $\textsf{R}$ than about $\textsf{B}$, or vice versa. However, it is safe to say that the above three expectations underlie the physics of duality. We also expect that in the de Sitter context, there are several key conflicts with some of these expectations, at least superficially. But on a more technical level, it is not out of the ordinary to expect that bulk reconstruction, subregion duality, operator algebras, and the black hole information aspects can be addressed as fundamentally as in AdS/CFT. Additionally, there are some interesting prospects of Jackiw-Teitelboim gravity in de Sitter space, drawing motivation from the highly successful works on JT gravity in AdS; see for instance some recent papers in de Sitter \cite{Balasubramanian:2020xqf,Moitra:2022glw,Nanda:2023wne,Aalsma:2021bit}.


		
{\bf Acknowledgements:}	We would like to thank Vasu Shyam. AV would like to thank the Indian Institute of Technology, Kanpur.

		\bibliography{references}

\providecommand{\href}[2]{#2}\begingroup\raggedright\begin{thebibliography}{10}

\bibitem{Maldacena:1997re}
J.~M. Maldacena, ``{The Large N limit of superconformal field theories and
  supergravity},'' \href{http://dx.doi.org/10.4310/ATMP.1998.v2.n2.a1}{{\em
  Adv. Theor. Math. Phys.} {\bfseries 2} (1998) 231--252},
  \href{http://arxiv.org/abs/hep-th/9711200}{{ arXiv:hep-th/9711200}}.

\bibitem{Witten:1998qj}
E.~Witten, ``{Anti-de Sitter space and holography},''
  \href{http://dx.doi.org/10.4310/ATMP.1998.v2.n2.a2}{{\em Adv. Theor. Math.
  Phys.} {\bfseries 2} (1998) 253--291},
  \href{http://arxiv.org/abs/hep-th/9802150}{{ arXiv:hep-th/9802150}}.

\bibitem{Aharony:1999ti}
O.~Aharony, S.~S. Gubser, J.~M. Maldacena, H.~Ooguri, and Y.~Oz, ``{Large N
  field theories, string theory and gravity},''
  \href{http://dx.doi.org/10.1016/S0370-1573(99)00083-6}{{\em Phys. Rept.}
  {\bfseries 323} (2000) 183--386},
  \href{http://arxiv.org/abs/hep-th/9905111}{{ arXiv:hep-th/9905111}}.

\bibitem{McGough:2016lol}
L.~McGough, M.~Mezei, and H.~Verlinde, ``{Moving the CFT into the bulk with $
  T\overline{T} $},'' \href{http://dx.doi.org/10.1007/JHEP04(2018)010}{{\em
  JHEP} {\bfseries 04} (2018) 010}, \href{http://arxiv.org/abs/1611.03470}{{
  arXiv:1611.03470 [hep-th]}}.

\bibitem{Shyam:2018sro}
V.~Shyam, ``{Finite Cutoff AdS$_{5}$ Holography and the Generalized Gradient
  Flow},'' \href{http://dx.doi.org/10.1007/JHEP12(2018)086}{{\em JHEP}
  {\bfseries 12} (2018) 086}, \href{http://arxiv.org/abs/1808.07760}{{
  arXiv:1808.07760 [hep-th]}}.

\bibitem{Araujo-Regado:2022gvw}
G.~Araujo-Regado, R.~Khan, and A.~C. Wall, ``{Cauchy slice holography: a new
  AdS/CFT dictionary},'' \href{http://dx.doi.org/10.1007/JHEP03(2023)026}{{\em
  JHEP} {\bfseries 03} (2023) 026}, \href{http://arxiv.org/abs/2204.00591}{{
  arXiv:2204.00591 [hep-th]}}.

\bibitem{Laddha:2020kvp}
A.~Laddha, S.~G. Prabhu, S.~Raju, and P.~Shrivastava, ``{The Holographic Nature
  of Null Infinity},''
  \href{http://dx.doi.org/10.21468/SciPostPhys.10.2.041}{{\em SciPost Phys.}
  {\bfseries 10} no.~2, (2021) 041}, \href{http://arxiv.org/abs/2002.02448}{{
  arXiv:2002.02448 [hep-th]}}.

\bibitem{deMelloKoch:2022sul}
R.~de~Mello~Koch and G.~Kemp, ``{Holography of information in AdS/CFT},''
  \href{http://dx.doi.org/10.1007/JHEP12(2022)095}{{\em JHEP} {\bfseries 12}
  (2022) 095}, \href{http://arxiv.org/abs/2210.11066}{{ arXiv:2210.11066
  [hep-th]}}.

\bibitem{Leutheusser:2021frk}
S.~A.~W. Leutheusser, ``{Emergent Times in Holographic Duality},''
  \href{http://dx.doi.org/10.1103/PhysRevD.108.086020}{{\em Phys. Rev. D}
  {\bfseries 108} no.~8, (2023) 086020},
  \href{http://arxiv.org/abs/2112.12156}{{ arXiv:2112.12156 [hep-th]}}.

\bibitem{Leutheusser:2022bgi}
S.~Leutheusser and H.~Liu, ``{Subalgebra-subregion duality: emergence of space
  and time in holography},'' \href{http://arxiv.org/abs/2212.13266}{{
  arXiv:2212.13266 [hep-th]}}.

\bibitem{Raju:2019qjq}
S.~Raju, ``{Is Holography Implicit in Canonical Gravity?},''
  \href{http://dx.doi.org/10.1142/S0218271819440115}{{\em Int. J. Mod. Phys. D}
  {\bfseries 28} no.~14, (2019) 1944011},
  \href{http://arxiv.org/abs/1903.11073}{{ arXiv:1903.11073 [gr-qc]}}.

\bibitem{Chakraborty:2023los}
T.~Chakraborty, J.~Chakravarty, V.~Godet, P.~Paul, and S.~Raju, ``{Holography
  of information in de Sitter space},''
  \href{http://dx.doi.org/10.1007/JHEP12(2023)120}{{\em JHEP} {\bfseries 12}
  (2023) 120}, \href{http://arxiv.org/abs/2303.16316}{{ arXiv:2303.16316
  [hep-th]}}.

\bibitem{dsnote}
A.~Verma and V.~Kalvakota, ``Revering Musings on de Sitter and Holography,
  2023,''. Available at \url{https://aayushayh.github.io/dSnote.pdf}.

\bibitem{Chakraborty:2023yed}
T.~Chakraborty, J.~Chakravarty, V.~Godet, P.~Paul, and S.~Raju, ``{The Hilbert
  space of de Sitter quantum gravity},''
  \href{http://dx.doi.org/10.1007/JHEP01(2024)132}{{\em JHEP} {\bfseries 01}
  (2024) 132}, \href{http://arxiv.org/abs/2303.16315}{{ arXiv:2303.16315
  [hep-th]}}.

\bibitem{moncrief1976space}
V.~Moncrief, ``Space--time symmetries and linearization stability of the
  Einstein equations. II,'' {\em Journal of Mathematical Physics} {\bfseries
  17} no.~10, (1976) 1893--1902.

\bibitem{higuchi1991quantum}
A.~Higuchi, ``Quantum linearization instabilities of de Sitter spacetime. I,''
  {\em Classical and Quantum Gravity} {\bfseries 8} no.~11, (1991) 1961.

\bibitem{higuchi1991quantum2}
A.~Higuchi, ``Quantum linearization instabilities of de Sitter spacetime. II,''
  {\em Classical and Quantum Gravity} {\bfseries 8} no.~11, (1991) 1983.

\bibitem{Ljatifi:2023dro}
M.~Ljatifi, ``{Group averaging and BRST quantization in de Sitter space},''
  \href{http://arxiv.org/abs/2305.11235}{{ arXiv:2305.11235 [hep-th]}}.

\bibitem{Xiao:2014uea}
X.~Xiao, ``{Holographic representation of local operators in de sitter
  space},'' \href{http://dx.doi.org/10.1103/PhysRevD.90.024061}{{\em Phys. Rev.
  D} {\bfseries 90} no.~2, (2014) 024061},
  \href{http://arxiv.org/abs/1402.7080}{{ arXiv:1402.7080 [hep-th]}}.

\bibitem{Jafferis:2015del}
D.~L. Jafferis, A.~Lewkowycz, J.~Maldacena, and S.~J. Suh, ``{Relative entropy
  equals bulk relative entropy},''
  \href{http://dx.doi.org/10.1007/JHEP06(2016)004}{{\em JHEP} {\bfseries 06}
  (2016) 004}, \href{http://arxiv.org/abs/1512.06431}{{ arXiv:1512.06431
  [hep-th]}}.

\bibitem{Faulkner:2017vdd}
T.~Faulkner and A.~Lewkowycz, ``{Bulk locality from modular flow},''
  \href{http://dx.doi.org/10.1007/JHEP07(2017)151}{{\em JHEP} {\bfseries 07}
  (2017) 151}, \href{http://arxiv.org/abs/1704.05464}{{ arXiv:1704.05464
  [hep-th]}}.

\bibitem{Dong:2016eik}
X.~Dong, D.~Harlow, and A.~C. Wall, ``{Reconstruction of Bulk Operators within
  the Entanglement Wedge in Gauge-Gravity Duality},''
  \href{http://dx.doi.org/10.1103/PhysRevLett.117.021601}{{\em Phys. Rev.
  Lett.} {\bfseries 117} no.~2, (2016) 021601},
  \href{http://arxiv.org/abs/1601.05416}{{ arXiv:1601.05416 [hep-th]}}.

\bibitem{Doi:2022iyj}
K.~Doi, J.~Harper, A.~Mollabashi, T.~Takayanagi, and Y.~Taki, ``{Pseudoentropy
  in dS/CFT and Timelike Entanglement Entropy},''
  \href{http://dx.doi.org/10.1103/PhysRevLett.130.031601}{{\em Phys. Rev.
  Lett.} {\bfseries 130} no.~3, (2023) 031601},
  \href{http://arxiv.org/abs/2210.09457}{{ arXiv:2210.09457 [hep-th]}}.

\bibitem{Doi:2023zaf}
K.~Doi, J.~Harper, A.~Mollabashi, T.~Takayanagi, and Y.~Taki, ``{Timelike
  entanglement entropy},''
  \href{http://dx.doi.org/10.1007/JHEP05(2023)052}{{\em JHEP} {\bfseries 05}
  (2023) 052}, \href{http://arxiv.org/abs/2302.11695}{{ arXiv:2302.11695
  [hep-th]}}.

\bibitem{Narayan:2020nsc}
K.~Narayan, ``{de Sitter future-past extremal surfaces and the entanglement
  wedge},'' \href{http://dx.doi.org/10.1103/PhysRevD.101.086014}{{\em Phys.
  Rev. D} {\bfseries 101} no.~8, (2020) 086014},
  \href{http://arxiv.org/abs/2002.11950}{{ arXiv:2002.11950 [hep-th]}}.

\bibitem{Narayan:2023zen}
K.~Narayan, ``{Further remarks on de Sitter space, extremal surfaces and time
  entanglement},'' \href{http://arxiv.org/abs/2310.00320}{{ arXiv:2310.00320
  [hep-th]}}.

\bibitem{Narayan:2017xca}
K.~Narayan, ``{On extremal surfaces and de Sitter entropy},''
  \href{http://dx.doi.org/10.1016/j.physletb.2018.02.010}{{\em Phys. Lett. B}
  {\bfseries 779} (2018) 214--222}, \href{http://arxiv.org/abs/1711.01107}{{
  arXiv:1711.01107 [hep-th]}}.

\bibitem{Coleman:2021nor}
E.~Coleman, E.~A. Mazenc, V.~Shyam, E.~Silverstein, R.~M. Soni, G.~Torroba, and
  S.~Yang, ``{De Sitter microstates from T$ \overline{T} $ +
  \ensuremath{\Lambda}$_{2}$ and the Hawking-Page transition},''
  \href{http://dx.doi.org/10.1007/JHEP07(2022)140}{{\em JHEP} {\bfseries 07}
  (2022) 140}, \href{http://arxiv.org/abs/2110.14670}{{ arXiv:2110.14670
  [hep-th]}}.

\bibitem{Lewkowycz:2019xse}
A.~Lewkowycz, J.~Liu, E.~Silverstein, and G.~Torroba, ``{$ T\overline{T} $ and
  EE, with implications for (A)dS subregion encodings},''
  \href{http://dx.doi.org/10.1007/JHEP04(2020)152}{{\em JHEP} {\bfseries 04}
  (2020) 152}, \href{http://arxiv.org/abs/1909.13808}{{ arXiv:1909.13808
  [hep-th]}}.

\bibitem{Batra:2024kjl}
G.~Batra, G.~B. De~Luca, E.~Silverstein, G.~Torroba, and S.~Yang, ``{Bulk-local
  dS$_3$ holography: the Matter with $T\bar T+\Lambda_2$},''
  \href{http://arxiv.org/abs/2403.01040}{{ arXiv:2403.01040 [hep-th]}}.

\bibitem{Shyam:2021ciy}
V.~Shyam, ``{$ \mathrm{T}\overline{\mathrm{T}} $ + \ensuremath{\Lambda}$_{2}$
  deformed CFT on the stretched dS$_{3}$ horizon},''
  \href{http://dx.doi.org/10.1007/JHEP04(2022)052}{{\em JHEP} {\bfseries 04}
  (2022) 052}, \href{http://arxiv.org/abs/2106.10227}{{ arXiv:2106.10227
  [hep-th]}}.

\bibitem{Franken:2023pni}
V.~Franken, H.~Partouche, F.~Rondeau, and N.~Toumbas, ``{Bridging the static
  patches: de Sitter holography and entanglement},''
  \href{http://dx.doi.org/10.1007/JHEP08(2023)074}{{\em JHEP} {\bfseries 08}
  (2023) 074}, \href{http://arxiv.org/abs/2305.12861}{{ arXiv:2305.12861
  [hep-th]}}.

\bibitem{Franken:2024ruw}
V.~Franken, ``{de Sitter Connectivity from Holographic Entanglement},'' in {\em
  {23rd Hellenic School and Workshops on Elementary Particle Physics and
  Gravity}}.
\newblock 3, 2024.
\newblock \href{http://arxiv.org/abs/2403.14889}{{ arXiv:2403.14889 [hep-th]}}.

\bibitem{Shaghoulian:2022fop}
E.~Shaghoulian and L.~Susskind, ``{Entanglement in De Sitter space},''
  \href{http://dx.doi.org/10.1007/JHEP08(2022)198}{{\em JHEP} {\bfseries 08}
  (2022) 198}, \href{http://arxiv.org/abs/2201.03603}{{ arXiv:2201.03603
  [hep-th]}}.

\bibitem{Chandrasekaran:2022eqq}
V.~Chandrasekaran, G.~Penington, and E.~Witten, ``{Large N algebras and
  generalized entropy},'' \href{http://dx.doi.org/10.1007/JHEP04(2023)009}{{\em
  JHEP} {\bfseries 04} (2023) 009}, \href{http://arxiv.org/abs/2209.10454}{{
  arXiv:2209.10454 [hep-th]}}.

\bibitem{Witten:2021unn}
E.~Witten, ``{Gravity and the crossed product},''
  \href{http://dx.doi.org/10.1007/JHEP10(2022)008}{{\em JHEP} {\bfseries 10}
  (2022) 008}, \href{http://arxiv.org/abs/2112.12828}{{ arXiv:2112.12828
  [hep-th]}}.

\bibitem{Engelhardt:2014gca}
N.~Engelhardt and A.~C. Wall, ``{Quantum Extremal Surfaces: Holographic
  Entanglement Entropy beyond the Classical Regime},''
  \href{http://dx.doi.org/10.1007/JHEP01(2015)073}{{\em JHEP} {\bfseries 01}
  (2015) 073}, \href{http://arxiv.org/abs/1408.3203}{{ arXiv:1408.3203
  [hep-th]}}.

\bibitem{Wall:2011hj}
A.~C. Wall, ``{A proof of the generalized second law for rapidly changing
  fields and arbitrary horizon slices},''
  \href{http://dx.doi.org/10.1103/PhysRevD.85.104049}{{\em Phys. Rev. D}
  {\bfseries 85} (2012) 104049}, \href{http://arxiv.org/abs/1105.3445}{{
  arXiv:1105.3445 [gr-qc]}}. [Erratum: Phys.Rev.D 87, 069904 (2013)].

\bibitem{Jensen:2023yxy}
K.~Jensen, J.~Sorce, and A.~J. Speranza, ``{Generalized entropy for general
  subregions in quantum gravity},''
  \href{http://dx.doi.org/10.1007/JHEP12(2023)020}{{\em JHEP} {\bfseries 12}
  (2023) 020}, \href{http://arxiv.org/abs/2306.01837}{{ arXiv:2306.01837
  [hep-th]}}.

\bibitem{Kudler-Flam:2023qfl}
J.~Kudler-Flam, S.~Leutheusser, and G.~Satishchandran, ``{Generalized Black
  Hole Entropy is von Neumann Entropy},''
  \href{http://arxiv.org/abs/2309.15897}{{ arXiv:2309.15897 [hep-th]}}.

\bibitem{AliAhmad:2023etg}
S.~Ali~Ahmad and R.~Jefferson, ``{Crossed product algebras and generalized
  entropy for subregions},'' \href{http://arxiv.org/abs/2306.07323}{{
  arXiv:2306.07323 [hep-th]}}.

\bibitem{Klinger:2023tgi}
M.~S. Klinger and R.~G. Leigh, ``{Crossed products, extended phase spaces and
  the resolution of entanglement singularities},''
  \href{http://dx.doi.org/10.1016/j.nuclphysb.2024.116453}{{\em Nucl. Phys. B}
  {\bfseries 999} (2024) 116453}, \href{http://arxiv.org/abs/2306.09314}{{
  arXiv:2306.09314 [hep-th]}}.

\bibitem{Chandrasekaran:2022cip}
V.~Chandrasekaran, R.~Longo, G.~Penington, and E.~Witten, ``{An algebra of
  observables for de Sitter space},''
  \href{http://dx.doi.org/10.1007/JHEP02(2023)082}{{\em JHEP} {\bfseries 02}
  (2023) 082}, \href{http://arxiv.org/abs/2206.10780}{{ arXiv:2206.10780
  [hep-th]}}.

\bibitem{Witten:2023qsv}
E.~Witten, ``{Algebras, Regions, and Observers},''
  \href{http://arxiv.org/abs/2303.02837}{{ arXiv:2303.02837 [hep-th]}}.

\bibitem{Bahiru:2023zlc}
E.~Bahiru, A.~Belin, K.~Papadodimas, G.~Sarosi, and N.~Vardian, ``{Holography
  and Localization of Information in Quantum Gravity},''
  \href{http://arxiv.org/abs/2301.08753}{{ arXiv:2301.08753 [hep-th]}}.

\bibitem{Raju:2021lwh}
S.~Raju, ``{Failure of the split property in gravity and the information
  paradox},'' \href{http://dx.doi.org/10.1088/1361-6382/ac482b}{{\em Class.
  Quant. Grav.} {\bfseries 39} no.~6, (2022) 064002},
  \href{http://arxiv.org/abs/2110.05470}{{ arXiv:2110.05470 [hep-th]}}.

\bibitem{Almheiri:2019hni}
A.~Almheiri, R.~Mahajan, J.~Maldacena, and Y.~Zhao, ``{The Page curve of
  Hawking radiation from semiclassical geometry},''
  \href{http://dx.doi.org/10.1007/JHEP03(2020)149}{{\em JHEP} {\bfseries 03}
  (2020) 149}, \href{http://arxiv.org/abs/1908.10996}{{ arXiv:1908.10996
  [hep-th]}}.

\bibitem{Penington:2019npb}
G.~Penington, ``{Entanglement Wedge Reconstruction and the Information
  Paradox},'' \href{http://dx.doi.org/10.1007/JHEP09(2020)002}{{\em JHEP}
  {\bfseries 09} (2020) 002}, \href{http://arxiv.org/abs/1905.08255}{{
  arXiv:1905.08255 [hep-th]}}.

\bibitem{Papadodimas:2012aq}
K.~Papadodimas and S.~Raju, ``{An Infalling Observer in AdS/CFT},''
  \href{http://dx.doi.org/10.1007/JHEP10(2013)212}{{\em JHEP} {\bfseries 10}
  (2013) 212}, \href{http://arxiv.org/abs/1211.6767}{{ arXiv:1211.6767
  [hep-th]}}.

\bibitem{Balasubramanian:2020xqf}
V.~Balasubramanian, A.~Kar, and T.~Ugajin, ``{Islands in de Sitter space},''
  \href{http://dx.doi.org/10.1007/JHEP02(2021)072}{{\em JHEP} {\bfseries 02}
  (2021) 072}, \href{http://arxiv.org/abs/2008.05275}{{ arXiv:2008.05275
  [hep-th]}}.

\bibitem{Moitra:2022glw}
U.~Moitra, S.~K. Sake, and S.~P. Trivedi, ``{Aspects of Jackiw-Teitelboim
  gravity in Anti-de Sitter and de Sitter spacetime},''
  \href{http://dx.doi.org/10.1007/JHEP06(2022)138}{{\em JHEP} {\bfseries 06}
  (2022) 138}, \href{http://arxiv.org/abs/2202.03130}{{ arXiv:2202.03130
  [hep-th]}}.

\bibitem{Nanda:2023wne}
K.~K. Nanda, S.~K. Sake, and S.~P. Trivedi, ``{JT gravity in de Sitter space
  and the problem of time},''
  \href{http://dx.doi.org/10.1007/JHEP02(2024)145}{{\em JHEP} {\bfseries 02}
  (2024) 145}, \href{http://arxiv.org/abs/2307.15900}{{ arXiv:2307.15900
  [hep-th]}}.

\bibitem{Aalsma:2021bit}
L.~Aalsma and W.~Sybesma, ``{The Price of Curiosity: Information Recovery in de
  Sitter Space},'' \href{http://dx.doi.org/10.1007/JHEP05(2021)291}{{\em JHEP}
  {\bfseries 05} (2021) 291}, \href{http://arxiv.org/abs/2104.00006}{{
  arXiv:2104.00006 [hep-th]}}.

\end{thebibliography}\endgroup
		\bibliographystyle{utphys}
	\end{document}